\documentclass[letterpaper, 10 pt, conference]{ieeeconf}  % Comment this line out if you need a4paper

\IEEEoverridecommandlockouts                              % This command is only needed if 
\usepackage{times} % assumes new font selection scheme installed
\usepackage{amsmath}
\usepackage{mathtools}
\usepackage{xcolor}
\usepackage{algorithm}
\usepackage{algpseudocode}
\usepackage{amssymb}  % assumes amsmath package installed
\usepackage{comment}
\usepackage{subcaption}
\usepackage{multirow}
\usepackage{graphicx}
\usepackage{booktabs}
\usepackage{tabularx}

\newcolumntype{C}{>{\centering\arraybackslash}X}

\title{\LARGE \bf
A Hybrid Particle Gaussian Mixture Filtering Method for Cislunar Target Tracking Under Extreme Uncertainty}

\author{Ishan Paranjape$^{1}$, Tarun Hejmadi$^{1}$, Utkarsh Ranjan Mishra$^{2}$, and Suman Chakravorty$^{3}$% <-this % stops a space
\thanks{$^{1}$Ishan Paranjape and Tarun Hejmadi are Graduate Research Assistants with the Department of Aerospace Engineering at Texas A\&M University, College Station, TX, USA
        {\tt\small {\{ishan.paranjape, tarunhejmadi}\}@tamu.edu}}%
\thanks{$^{2}$Utkarsh Ranjan Mishra is a Senior Sensor Fusion Engineer at Lucid Group, Inc.
        {\tt\small utkarshmishra@lucidmotors.com}}%
\thanks{$^{3}$Suman Chakravorty a Professor with the Department of Aerospace Engineering, Texas A\&M University, College Station, TX, USA
        {\tt\small schakrav@tamu.edu}}%
}

\begin{document}

\maketitle
\thispagestyle{empty}
\pagestyle{empty}

%%%%%%%%%%%%%%%%%%%%%%%%%%%%%%%%%%%%%%%%%%%%%%%%%%%%%%%%%%%%%%%%%%%%%%%%%%%%%%%%
\begin{abstract}

Gauss's method of orbit determination (OD) and its variants are among the most popular initial state estimation techniques for astronomers and engineers alike. However, owing to its assumptions regarding the two-body problem, Gauss's method is inapplicable in the cislunar domain, where three body effects dominate. We introduce a hybrid Particle Gaussian Mixture filtering method, a purely recursive probabilistic orbit determination framework based on a combination of the Markov Chain Monte Carlo based Particle Gaussian Mixture-II (PGM-II) and Particle Gaussian Mixture-I (PGM-I) filters. This method enables us to fuse probabilistic information with angles-only observations from terrestrial telescopes for short and long-term cislunar target tracking. We demonstrate this technique on an important cislunar orbit regime. 

\end{abstract}

\section{INTRODUCTION}

Every probabilistic target tracking problem associated with outer space begins with the generation of an initial state estimate, also known as initial orbit determination (IOD). To this day, Gauss's method of orbit determination (OD) -- consisting of IOD and nonlinear least squares refinement of the IOD estimate -- remains one of the most powerful methods for generating a deterministic initial state estimate leveraging three consecutive, closely-spaced angle observations \cite{taff1979}. Gauss's method's probabilistic counterpart, as well as more specialized IOD methods such as probabilistic admissible region (PAR) persist throughout space situational awareness (SSA) literature, especially for regimes defined by the two-body problem \cite{demars2013, hussein2014, mishra2024}. In this article, we focus on probabilistic target tracking in the cislunar domain -- defined by the three-body problem with the Earth and Moon as primaries -- where Gauss's method is no longer valid \cite{holzinger2021}. The literature on such problems is sparse, and includes PAR-based, sparse grid collocation, and kinematic fitting methods \cite{bolden2022, heidrich2025, paranjape2026, paranjape2025}. 

After generating an initial state estimate, subsequent orbit determination typically requires filtering. The Kalman Filter is typically used for linear systems with Gaussian \textit{a priori} estimates and sensor noise \cite{kalmanBucy1961}. For nonlinear systems, variants such as the Extended Kalman Filter (EKF), Unscented Kalman Filter (UKF), and Ensemble Kalman Filter (EnKF) are used \cite{crassidis2011, wan2000, evensen2003}. In the cislunar and near-Earth regimes, the batch least squares filter is a popular OD algorithm \cite{tapley2004}. Since such systems have trajectories that can behave chaotically with time, Gaussian mixture model (GMM) based filters such as adaptive GMM filters, ensemble GMM filters, and UKF/particle hybrid filters have been developed \cite{popov2023, durant2023, iannamorelli2025, raihan2017hybrid}. Ref. \cite{raihan2017hybrid} in particular spurred the development of the Particle Gaussian Mixture (PGM) filters \cite{raihan2018pgm1, raihan2018pgm2}.

The Particle Gaussian Mixture filters are a set of related techniques which utilize particle-based estimate propagation and clustering-based interactive mixture modeling. The first PGM filter, PGM-I, addresses the issues of inflexibility in the number of GMM components between filtering steps and particle depletion that arises from particle filters' update step. This filter has been successfully demonstrated for numerous chaotic dynamical systems, including two-body and three-body systems in space \cite{mishra2024, bolden2022, paranjape2025, paranjape2026, givens2025}. The second PGM filter, PGM-II, was developed to address situations in which a Kalman update is insufficient or where multi-modality is expected in the update step. This filter also addresses inflexibility in the number of GMM components between the propagation and update steps of a single filtering iteration \cite{raihan2018pgm2}. In its update process, the PGM-II filter utilizes Markov Chain Monte Carlo (MCMC) sampling, along with the GMM description of the predicted prior and the measurement likelihood function, to sample from the posterior. %This article focuses on an extension and application of this PGM-II update step.

In a cislunar environment, the framework in Refs. \cite{paranjape2026} and \cite{paranjape2025} -- hereafter referred to as the Kinematic Fitting Particle Gaussian Mixture (KF-PGM) framework -- presents a simple IOD technique using kinematic fitting and a few assumptions about the target's states. When combined with the PGM-I Filter, this technique robustly tracks targets in a cislunar environment for long periods of time even in the event of long sensor shutoffs. However, the kinematic fitting IOD algorithm requires curve fitting through several initial observations (8-10) with relatively high order fits. In reality, we may only be able to make a handful of observations of the target in a single day or week, requiring us to leverage as few observations of the target state as possible much like Gauss's method.

%Although the KF-PGM makes minimal assumptions about a target's characteristics, we may obtain more reliable information about the target we are tracking (e.g. orbit type, range bounds, and angular rates). Probabilistic fusion of this information with the KF-PGM framework is only possible if such information is distributed as zero-mean Gaussian given the underlying target state. 
In this paper, we introduce a purely recursive approach, unlike existing cislunar IOD/OD approaches like KF-PGM, involving both Particle Gaussian Mixture filters for cislunar target tracking. The only \textit{a priori} information that this method requires is that the object belongs to the cislunar domain, and angles-only observations from terrestrial telescopes. We utilize this information to accurately localize objects within a single pass when the object is visible to the observer. This method utilizes a hybrid of the PGM-I and PGM-II based filtering over the same measurement sets as KF-PGM to create up to a 100-fold improvement in position and velocity estimates. Thus, this approach provides an accurate and data-efficient recursive solution to the problem of target tracking in cislunar space.

The remainder of this article is organized as follows. First, we set up the problem of cislunar target tracking by providing context about cislunar dynamics and the measurement model. Next, we provide some mathematical and algorithmic background about the PGM-I and PGM-II filters, and how our hybrid PGM filtering method uses a modified PGM-II update step and uses both filters in sequence. Finally, we demonstrate this hybrid PGM filter for a portion of a popular cislunar orbit and compare its effectiveness with that of the KF-PGM framework.

\section{Cislunar Dynamics and Observation Model}\label{sec:cislunar}
In this section, we describe the dynamics and measurements relevant to cislunar target tracking. The cislunar domain is defined as the full $4\pi$ steradian volume in which three-body dynamics including the Earth and Moon as primaries are modeled \cite{bolden2022}. Interest in this region has burgeoned with the rise of lunar missions such as China's Chang-e' 5 or India's Chandrayaan. Furthermore, the United States has outlined effective space situational awareness in this domain as a matter of national importance \cite{ussf2022}. %For this reason, we focus on applying our hybrid PGM-based filtering approach to this domain.

Cislunar dynamics are defined by the three-body problem. The most popular dynamics model in literature is the circular-restricted three-body problem (CR3BP), in which the two larger bodies -- the Earth and the Moon -- orbit around their center of mass, and a third body -- the target or resident space object (RSO) of interest -- moves relative to the rotating motion of the two primary bodies. In this barycentric or synodic reference frame, state dynamics for $\mathbf{x} = [x, y, z, \dot{x}, \dot{y}, \dot{z}]^T$ are modeled as follows:
\begin{subequations} \label{eq18:cr3bp}
    \begin{align}
        \ddot{x} = x + 2\dot{y} - \frac{(1-\mu)(x+\mu)}{r_1^3} - \frac{\mu x - \mu(1-\mu)}{r_2^3}\label{eq18a:cr3bp_X} \\
        \ddot{y} = y - 2 \dot{x} - \frac{(1-\mu)y}{r_1^3} - \frac{\mu y}{r_2^3}\label{eq18b:cr3bp_Y} \\
        \ddot{z} = \frac{(1-\mu)z}{r_1^3} - \frac{\mu z}{r_2^3} \label{eq18c:cr3bp_Z}
    \end{align}
\end{subequations}
Physical units are normalized per Ref. \cite{schaub2003}. Furthermore, $\mu$ represents the mass ratio of the two largest bodies, and $r_1$ and $r_2$ represent the distance between the target and the Earth and Moon, respectively. 

Measurements of cislunar RSOs are taken from ground-based optical telescopes, which have the ability to measure azimuth and elevation angles. Letting $\mathbf{x}^{\mathcal{T}} = [x^{\mathcal{T}}, y^{\mathcal{T}}, z^{\mathcal{T}}, \dot{x}^{\mathcal{T}}, \dot{y}^{\mathcal{T}}, \dot{z}^{\mathcal{T}}]^T$ represent the state vector $\mathbf{x}$ in the observer-centered topocentric frame, we define our measurement model $\mathbf{z}_k = [AZ, EL]^T$ as follows:
\begin{subequations} \label{eq19:AZ-EL}
    \begin{align}
        AZ = tan^{-1} \frac{y^{\mathcal{T}}}{x^{\mathcal{T}}}\label{eq19a:azimuth} \\
        EL = \frac{\pi}{2} - cos^{-1} \frac{z^{\mathcal{T}}}{\sqrt{(x^{\mathcal{T}})^2 + (y^{\mathcal{T}})^2 + (z^{\mathcal{T}})^2}} \label{eq19b:elevation}
    \end{align}
\end{subequations}
For computational efficiency, we choose to represent our state vector $\mathbf{x}$ exclusively in this observer-centered topocentric frame, based in College Station, TX, USA. Conversions between the CR3BP synodic frame and this topocentric reference frame are detailed in Ref. \cite{paranjape2026}.

\section{Particle Gaussian Mixture Filters}\label{sec:pgmfs}

As our article focuses on the application of a hybrid PGM filtering approach in a cislunar environment, this section outlines the general PGM-I and PGM-II approaches.

\subsection{Bayesian Filtering Preliminaries}\label{subsec:bayesFilter}
The $n_x$-dimensional state of an object is represented as $\mathbf{x} \in \mathbb{R}^{n_x}$. The object state is propagated using a Markov chain whose transition density is represented by the PDF $p(\mathbf{x}_{k}|\mathbf{x}_{k-1})$ at some time steps $k$ and $k-1$. An $n_z$-dimensional observation vector $\mathbf{z} \in \mathbb{R}^{n_z}$ is related to the object state by an observation function (or measurement model) $\mathbf{h}(\mathbf{x})$ with some measurement noise. This measurement noise is modeled as additive, zero-mean Gaussian and defined by the conditional PDF $p(\mathbf{z}_k|\mathbf{x})$. Let $p(\mathbf{x}|\mathbf{Z}^{k-1}) = p_{k-1}^{+}(\mathbf{x})$ describe the conditional PDF for $\mathbf{x}$ given the sequence of observations $\mathbf{Z}^{k-1}$ up to time step $k-1$, or as the posterior estimate for $\mathbf{x}$ at time $k-1$. Using the Markov transition density function, the \textit{a priori} probability density of $\mathbf{x}$ for the current time step is defined as:
\begin{equation} \label{eq1:bayesProp}
    p_{k}^{-}(\mathbf{x}) = p_{k-1}^{+}(\mathbf{x}) p(\mathbf{x}_k|\mathbf{x}_{k-1}).
\end{equation}

To obtain the posterior estimate at time step $k$, we apply Bayes' rule:
\begin{equation} \label{eq2:bayesRule}
    p_{k}^{+}(\mathbf{x}) = \frac{p_k^{-}(\mathbf{x}) p(\mathbf{z}_k|\mathbf{x})}{p(\mathbf{z}_k)} = \frac{p_k^{-}(\mathbf{x}) p(\mathbf{z}_k|\mathbf{x})}{\int p_k^{-}(\mathbf{x}) p(\mathbf{z}_k|\mathbf{x}) d\mathbf{x}}.
\end{equation}
 The Particle Gaussian Mixture filters assume that $p_{1 \cdots k}^{-}(\mathbf{x})$ may be expressed as some GMM $p_k^{-}(\mathbf{x}) = \sum_{j=1}^{N} \omega_j^{-} \mathcal{N}(\mathbf{x}; \mathbf{\mu}_j^{-}, \mathbf{P}_j^{-})$, where $\mathcal{N}(\cdot)$ represents a Gaussian PDF with component mean $\mathbf{\mu}_j^{-}$, component covariance $\mathbf{P}_j^{-}$, and weight $\omega_j^{-}$. With this assumption, we can show that the posterior PDF may also be expressed as a GMM. 

We start by expanding the fraction in Eq. \eqref{eq2:bayesRule} as follows:
\begin{equation}\label{eq3:bayesExpansion}
    \begin{aligned}
        p_k^{+}(\mathbf{x}) = \frac{p_k^{-}(\mathbf{x}) p(\mathbf{z}_k|\mathbf{x})}{\int p_k^{-}(\mathbf{x}) p(\mathbf{z}_k|\mathbf{x}) d\mathbf{x}} \\
        = \frac{\sum_{i=1}^{N} \omega_i^{-} \mathcal{N}(\mathbf{x}; \mathbf{\mu}_i^{-}, \mathbf{P}_i^{-}) p(\mathbf{z}_k|\mathbf{x})}{\int \sum_{j=1}^{N} \omega_j^{-} \mathcal{N}(\mathbf{x}; \mathbf{\mu}_j^{-}, \mathbf{P}_j^{-}) p(\mathbf{z}_k|\mathbf{x})} \\
        = \frac{\sum_{i=1}^{N} \omega_i^{-} \mathcal{N}(\mathbf{x}; \mathbf{\mu}_i^{-}, \mathbf{P}_i^{-}) p(\mathbf{z}_k|\mathbf{x})}{\sum_{j=1}^{N} \omega_j^{-} \int \mathcal{N}(\mathbf{x}; \mathbf{\mu}_i^{-}, \mathbf{P}_i^{-}) p(\mathbf{z}_k|\mathbf{x})}.
    \end{aligned}
\end{equation}
At this point, we define the likelihood function $l_i(\mathbf{z}_k) = \int \mathcal{N}(\mathbf{x}; \mathbf{\mu}_i^{-}, \mathbf{P}_i^{-}) p(\mathbf{z}_k|\mathbf{x})$. Substituting the result in Eq. \eqref{eq3:bayesExpansion} and multiplying the numerator and denominator by $l_i(\mathbf{z}_k)$, 
\begin{equation}\label{eq4:posteriorGMM}
    \begin{aligned}
        p_k^{+}(\mathbf{x}) = \frac{\sum_{i=1}^{N} \omega_i^{-} \mathcal{N}(\mathbf{x}; \mathbf{\mu}_i^{-}, \mathbf{P}_i^{-}) p(\mathbf{z}_k|\mathbf{x})}{\sum_{j=1}^{N} \omega_j^{-} l_i(\mathbf{z}_k)} \frac{l_i(\mathbf{z}_k)}{l_i(\mathbf{z}_k)} \\
        = \sum_{i=1}^{N} \frac{\omega_i^{-} l_i(\mathbf{z}_k)}{\sum_{j=1}^{N} \omega_j^{-} l_i(\mathbf{z}_k)} \frac{\mathcal{N}(\mathbf{x}; \mathbf{\mu}_i^{-}, \mathbf{P}_i^{-}) p(\mathbf{z}_k|\mathbf{x})}{l_i(\mathbf{z}_k)} \\
        = \sum_{i=1}^{N} \omega_i^{+} p_{i,k}^{+}(\mathbf{x}).
    \end{aligned}
\end{equation}

The expansion in Eq. \eqref{eq4:posteriorGMM} shows that the Bayesian update for a GMM can be split into a discrete weight update and a continuous PDF update, respectively:
\begin{subequations}\label{eq5:posteriorUpdates}
    \begin{align}
        \omega_i^{+} = \frac{\omega_i^{-}l_i(\mathbf{z_k})}{\sum_{j=1}^{N} \omega_j^{-}l_j(\mathbf{z}_k)} \label{eq5a:discreteUpdate} \\ 
        p_{i,k}^{+}(\mathbf{x}) = \frac{\mathcal{N}(\mathbf{x}; \mathbf{\mu}_i^{-}, \mathbf{P}_i^{-}) p(\mathbf{z}_k|\mathbf{x})}{l_i(\mathbf{z}_k)} \label{eq5b:continuousUpdate} . 
    \end{align}
\end{subequations}
While the discrete (i.e. weight) update in Eq. \eqref{eq5a:discreteUpdate} is consistent between the two PGM filters, the likelihood computation $l_i(\mathbf{z}_k)$ and continuous update in Eq. \eqref{eq5b:continuousUpdate} are not. These differences are described in greater detail in the proceeding subsections.

Between the particle propagation and update steps, both PGM filters require an \textit{a priori} estimate modeled as a GMM. Propagated particles may be approximated or fitted into a GMM using some clustering algorithm $\mathcal{C}$, such as \textit{k-means++}, such that the set of particles is partitioned into $N$ different sets or ensembles, each with an associated weight, mean, and covariance \cite{lloyd1982}. The PGM filters bypass particle depletion by performing Bayesian updates on each GMM component rather than the particles of the \textit{a priori} estimate. While the PGM-I filter uses this clustering step only once, the PGM-II filter utilizes this clustering step twice. Furthermore, both PGM filters rely on resampling several particles from the posterior estimate (also modeled as a GMM) prior to propagating the ensemble of particles to the next time step.%\\
%\textit{The key to avoiding particle depletion in the PGM is to find a GMM approximation of the predicted ensemble and performing the Bayes update on the GMM rather than the particles themselves as in a particle filter.} 

\subsection{Particle Gaussian Mixture-I Filter}\label{subsec:pgm1}
The major differences between the two PGM filters lie during the Bayesian update process. For the PGM-I filter specifically, the continuous update involves a Kalman update, such that
\begin{equation}
    p_{i,k}^{+}(\mathbf{x}) = \frac{\mathcal{N}(\mathbf{x}; \mathbf{\mu}_i^{-}, \mathbf{P}_i^{-}) p(\mathbf{z}_k|\mathbf{x})}{l_i(\mathbf{z}_k)} = \mathcal{N}(\mathbf{x}; \mathbf{\mu}_i^{+}, \mathbf{P}_i^{+}).
\end{equation}
While EKF and UKF-based updates are possible, it is highly recommended to utilize a particle-based ensemble (i.e. EnKF) update for the mean and covariance as outlined in Refs. \cite{raihan2018pgm1} and \cite{mishra2024}. A single iteration of the PGM-I filter is demonstrated by Refs. \cite{paranjape2026} and \cite{paranjape2025}, and the PGM-I algorithm (copied from Ref. \cite{raihan2018pgm1}) is provided below.

\begin{algorithm}
\caption{Particle Gaussian Mixture Filter I}\label{alg:pgm1}
\begin{algorithmic}
    \State Given $\pi_0 (x) = \sum_{i=1}^{M(0)} \omega_i (0) p_g(x; \mu_i (0), P_i (0))$, transition density kernel $p(x'|x)$, $n = 1$.
    \begin{enumerate}
        \State Sample $N_p$ particles from $\pi_{n-1}$ and the transition density kernel $p_n (x'|x)$ as follows: \label{alg:repeat}
        \begin{itemize}
            \item Sample $X^{(i)^\prime}$ from $\pi_{n-1} (.)$.
            \item Sample $X^{(i)}$ from $p(.|X^{(i)^\prime})$.
        \end{itemize}
        \State Use a clustering algorithm $\mathcal{C}$ to cluster the set of particles $X^{(i)}$ into $M^{-} (n)$ Gaussian clusters with weights, mean, and covariance given by $\{\omega_i^{-} (n), \mu_i^{-} (n), P_i^{-} (n)\}$.
        \State Update the mixture weights and mixture means and covariances to $\{\omega_i^{+} (n), \mu_i^{+} (n), P_i^{+} (n)\}$, given the observation $z_n$, utilizing the Kalman update outlined in Refs. \cite{raihan2018pgm1} and \cite{mishra2024}.
        \State $n = n + 1$. Go to Step \ref{alg:repeat}.
    \end{enumerate}
\end{algorithmic}\end{algorithm}

\subsection{Particle Gaussian Mixture-II Filter}\label{subsec:pgm2}
The propagation and clustering steps of the PGM-I and PGM-II filters (i.e. Steps 1 and 2 of Algorithm \ref{alg:pgm1}) are identical. However, since the PGM-II filter was designed to handle cases in which a Kalman or Kalman-family update would fail to reduce any uncertainty or fail to capture the nature of the resulting PDF, the PGM-II filter's update process consists of three major parts: 1) MCMC sampling, 2) clustering MCMC samples, and 3) weight update. 

The first step of the update process -- MCMC sampling -- samples from the posterior PDF $p_k^{+}(\mathbf{x}) \propto p_k^{-}(\mathbf{x})p(\mathbf{z}_k|\mathbf{x})$. A popular MCMC algorithm for PGM-II is the Metropolis algorithm \cite{hastings1970}. 
 An important component of this Metropolis algorithm is the development of a target function or PDF. We can mathematically define this target or sampling function $P(\mathbf{x})$ (denoted as the "posterior" in Ref. \cite{raihan2018pgm2}) as
\begin{equation}\label{eq9:MCMCsamplingfunc}
    P(\mathbf{x}) = p_k^{-}(\mathbf{x})p(\mathbf{z}_k|\mathbf{x}).
\end{equation}
We denote the sampling function with a capital $P$ since it is not a proper PDF, but is only proportional to the posterior. 

Since $p_k^{-}(\mathbf{x})$ is a GMM, using $P(\mathbf{x}) = \sum_{i=1}^{N} \omega_i^{-} \mathcal{N}(\mathbf{x}; \mathbf{\mu}_i^{-}, \mathbf{P}_i^{-})p(\mathbf{z}_k|\mathbf{x})$ becomes computationally intensive. For this reason, Ref. \cite{raihan2018pgm2} recommends MCMC sampling component-by-component. In mathematical terms, $P_i(\mathbf{x}) = p_{i,k}^{-}(\mathbf{x})p(\mathbf{z}_k|\mathbf{x})$. The next step of the Metropolis-based sampling step is to choose some symmetric proposal PDF $q(\mathbf{x})$ which allows the MCMC chain to explore the posterior PDF. For sufficient and computationally efficient sampling in the state space, multiple Markov chains must be run in parallel, each initialized differently. All resulting MCMC chains sampling target distribution $P_i(\mathbf{x})$ create an ensemble $A_i$.

The second part of the PGM-II update process involves clustering samples resulting from the first part. MCMC chains can converge to multiple local minima depending on their starting location. For this reason, it is important to discard any segments of the Markov chain during a certain burn-in period. Moreover, due to MCMC random walk behavior, samples in each chain need to be de-correlated. This is accomplished by only sampling every $S$-th sample/state in each Markov chain, where $S$ is a large number, typically in the high hundreds or thousands. Correlations are further removed by clustering remaining samples of $A_i$ in a similar fashion as the PGM-I clustering step. The result of this clustering is a functional representation $p_i^{+}(\mathbf{x})$ of the ensemble expressed as a GMM $p_i^{+}(\mathbf{x}) = \sum_{m=1}^{N_i} \omega_m^{*} \mathcal{N}(\mathbf{x}; \mathbf{\mu}_m^{+}, \mathbf{P}_m^{+})$, where $N_i$ represents the number of clusters. Different ensembles $A_i$ may have different $N_i$, depending on the shape of the resulting MCMC chains. The weights $\omega_m^{*}$ will be multiplied by the ensemble posterior weights $\omega_i^{+}$.

The final part of the PGM-II update process concerns the weight update and computation of ensemble likelihoods. As mentioned in Section \ref{subsec:pgm1}, the ensemble weight update formula is given by Eq. \eqref{eq5a:discreteUpdate}. However, likelihood computation differs. We derive the PGM-II likelihood approximation by considering some proper PDF $\pi(\mathbf{x})$. For each ensemble $i$, we start with the continuous PDF update given by Eq. \eqref{eq5b:continuousUpdate}.
\begin{equation}\label{eq11:cuExpansion1}
        p_{i,k}^{+}(\mathbf{x}) = \frac{\mathcal{N}(\mathbf{x}; \mathbf{\mu}_i^{-}, \mathbf{P}_i^{-}) p(\mathbf{z}_k|\mathbf{x})}{l_i(\mathbf{z}_k)} \frac{\pi(\mathbf{x})}{\pi(\mathbf{x})}.
\end{equation}
Rearranging terms, we obtain
\begin{equation}\label{eq12:cuExpansion2}
        \frac{l_{i}(\mathbf{z}_k)}{\pi(\mathbf{x})} = \frac{\mathcal{N}(\mathbf{x}; \mathbf{\mu}_i^{-}, \mathbf{P}_i^{-}) p(\mathbf{z}_k|\mathbf{x})}{p_{i,k}^{+}(\mathbf{x}) \pi(\mathbf{x})}.
\end{equation}
Taking the inverse of both sides,
\begin{equation}\label{eq13:cuExpansion3}
        \frac{\pi(\mathbf{x})}{l_{i}(\mathbf{z}_k)} = \frac{p_{i,k}^{+}(\mathbf{x}) \pi(\mathbf{x})}{\mathcal{N}(\mathbf{x}; \mathbf{\mu}_i^{-}, \mathbf{P}_i^{-}) p(\mathbf{z}_k|\mathbf{x})}.
\end{equation}
Taking the integral of both sides with respect to $d\mathbf{x}$,
\begin{equation}\label{eq14:cuExpansion4}
    \begin{aligned}
        \int_{\mathbb{R}^{n_x}} \frac{\pi(\mathbf{x})}{l_{i}(\mathbf{z}_k)} d\mathbf{x} = \int_{\mathbb{R}^{n_x}} \frac{p_{i,k}^{+}(\mathbf{x}) \pi(\mathbf{x})}{\mathcal{N}(\mathbf{x}; \mathbf{\mu}_i^{-}, \mathbf{P}_i^{-}) p(\mathbf{z}_k|\mathbf{x})} d\mathbf{x} \\
        = \frac{1}{l_i(\mathbf{z}_k)} = \int_{\mathbb{R}^{n_x}} \frac{p_{i,k}^{+}(\mathbf{x}) \pi(\mathbf{x})}{\mathcal{N}(\mathbf{x}; \mathbf{\mu}_i^{-}, \mathbf{P}_i^{-}) p(\mathbf{z}_k|\mathbf{x})} d\mathbf{x}.
    \end{aligned}
\end{equation}

Recognizing the discrete nature of each ensemble $A_i$, we can compute the ensemble likelihood $l_i(\mathbf{z}_k)$ by recognizing the following property:
\begin{equation}\label{eq15:discreteLikelihood}
    \begin{aligned}
        \frac{1}{l_i(\mathbf{z}_k)} = \int_{\mathbb{R}^{n_x}} \frac{p_{i,k}^{+}(\mathbf{x}) \pi(\mathbf{x})}{\mathcal{N}(\mathbf{x}; \mathbf{\mu}_i^{-}, \mathbf{P}_i^{-}) p(\mathbf{z}_k|\mathbf{x})} d\mathbf{x} \\
        = \mathbb{E}\left[ \frac{\pi(\mathbf{x})}{p(\mathbf{z}_k|\mathbf{x}) \mathcal{N}(\mathbf{x}; \mathbf{\mu}_i^{-}, \mathbf{P}_i^{-})} \right] \\
        \approx \frac{1}{S_i} \sum_{s=1}^{S_i} \frac{\pi(\mathbf{x}_s)}{p(\mathbf{z}_k|\mathbf{x}_s) \mathcal{N}(\mathbf{x}_s; \mathbf{\mu}_i^{-}, \mathbf{P}_i^{-})}.
    \end{aligned}
\end{equation}
In this approximation of the likelihood that $\mathbf{z}_k$ is explained by the $i$-th MCMC-based ensemble, $\mathbf{x}_s$ refers to each sample within $A_i$ (after removing burn-in samples and decorrelating samples). While any proper PDF $\pi(\mathbf{x}_s)$ holds, likelihood computation is further simplified by setting $\pi(\mathbf{x}_s) = \mathcal{N}(\mathbf{x}_s; \mathbf{\mu}_i^{-}, \mathbf{P}_{i}^{-})$. Once all ensemble likelihoods are computed, the resulting posterior ensemble weights $\omega_i^{+}$ are multiplied by the ensemble-clustered weights $\omega_m^{*}$ to obtain the posterior weights for each posterior component-tuple $(i,m)$. Numerous particles are drawn from this resulting GMM and propagated, leading to the next iteration of the PGM-II filter.

\begin{figure}[!thpb]
    \centering
    % Subfigure 1: Width of subfigure matches the image width
    \begin{subfigure}{0.8\columnwidth}
        \centering
        \includegraphics[width=\linewidth]{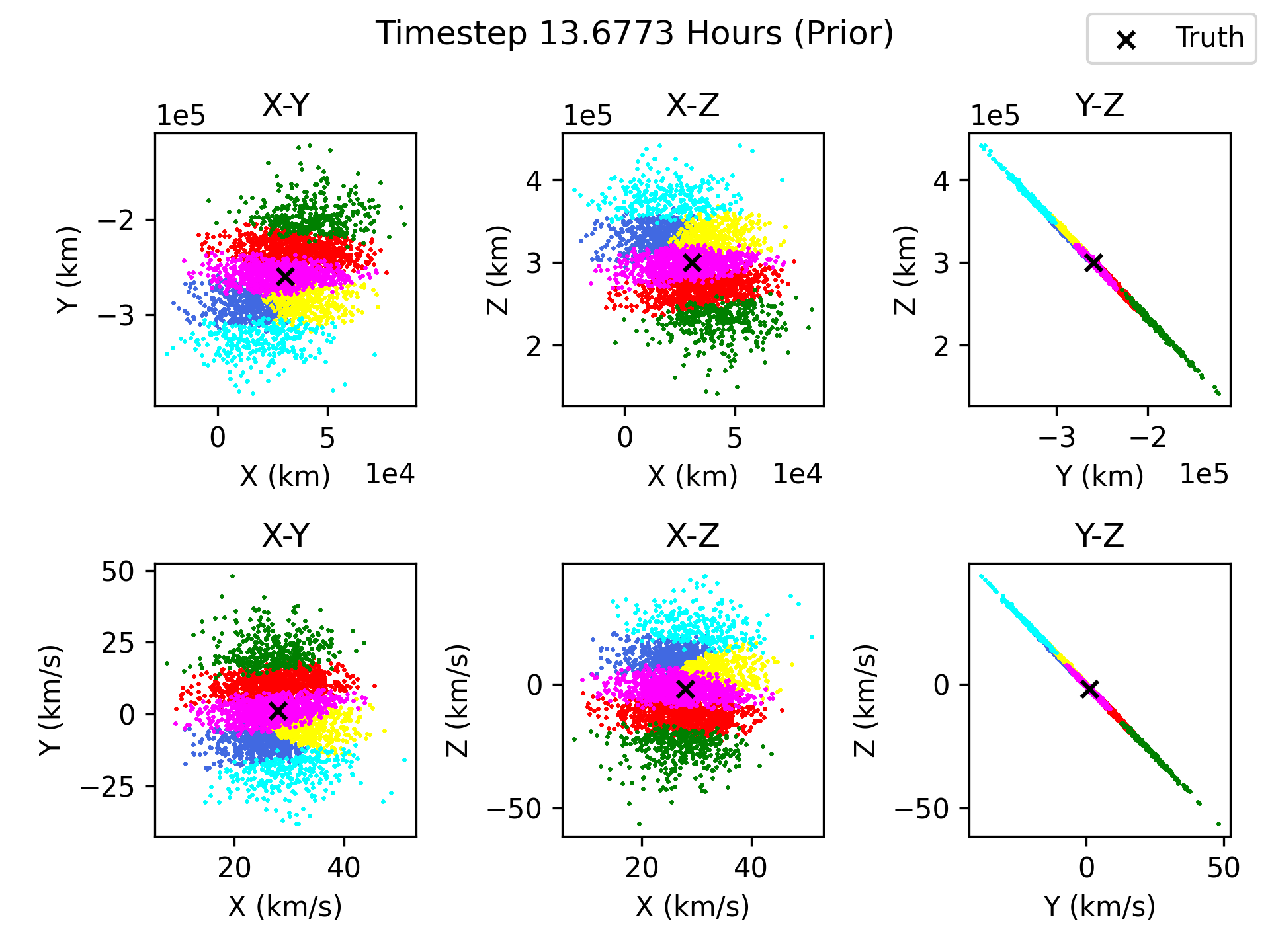}
        \caption{\textit{A priori} clustered estimate}
        \label{fig:2apgm2apriori}
    \end{subfigure}

    \vspace{0.2cm} % Optional vertical spacing

    % Subfigure 2: Width of subfigure matches the image width
    \begin{subfigure}{0.8\columnwidth}
        \centering
        \includegraphics[width=\linewidth]{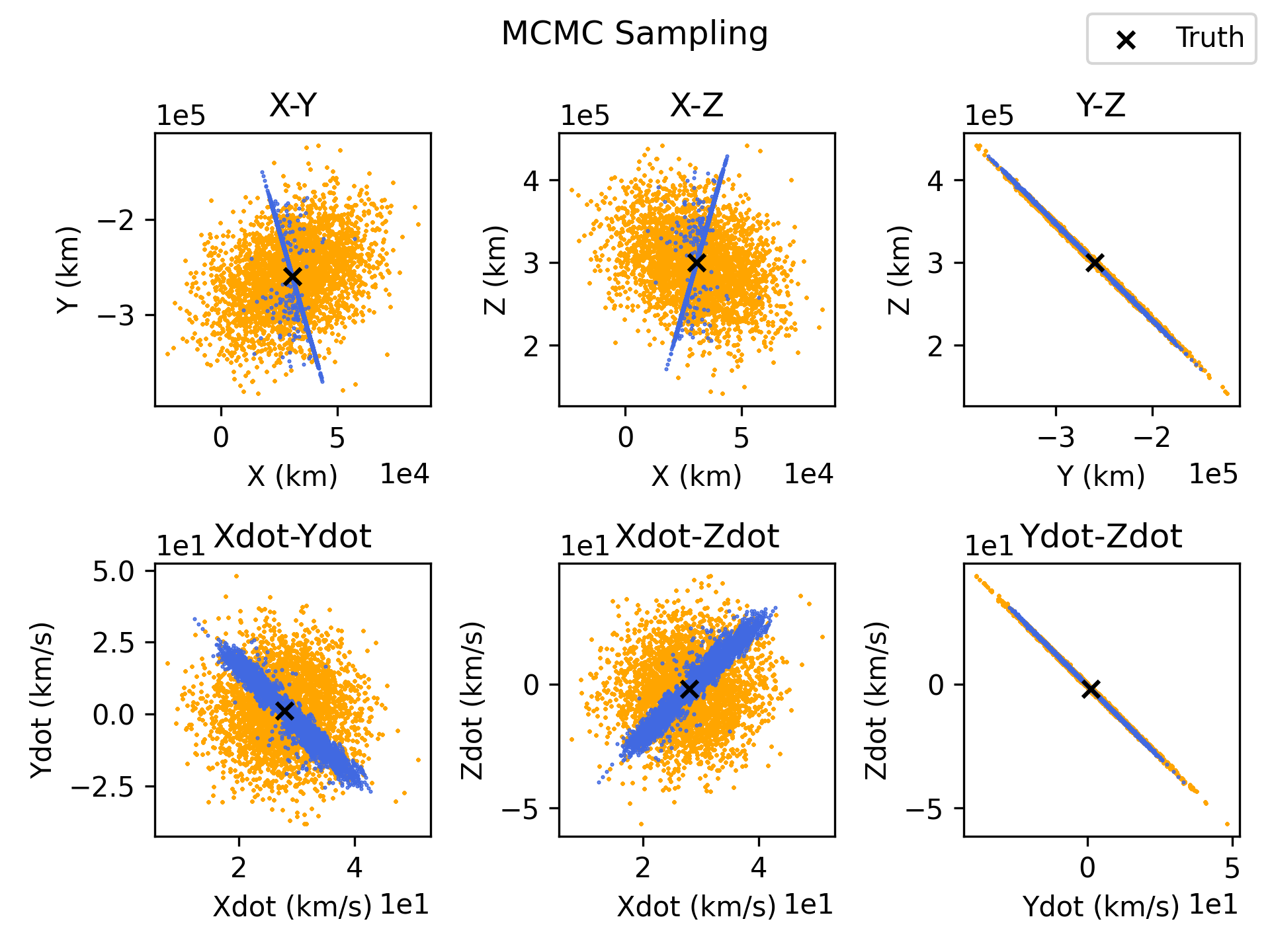}
        \caption{Collective MCMC chains $A_i$ for all $i$}
        \label{fig:2bpgm2mcmcAll}
    \end{subfigure}

    \vspace{0.2cm} % Optional vertical spacing

    %\begin{comment}
    % Subfigure 3: Width of subfigure matches the image width
    \begin{subfigure}{0.81\columnwidth}
        \centering
        \includegraphics[width=\linewidth]{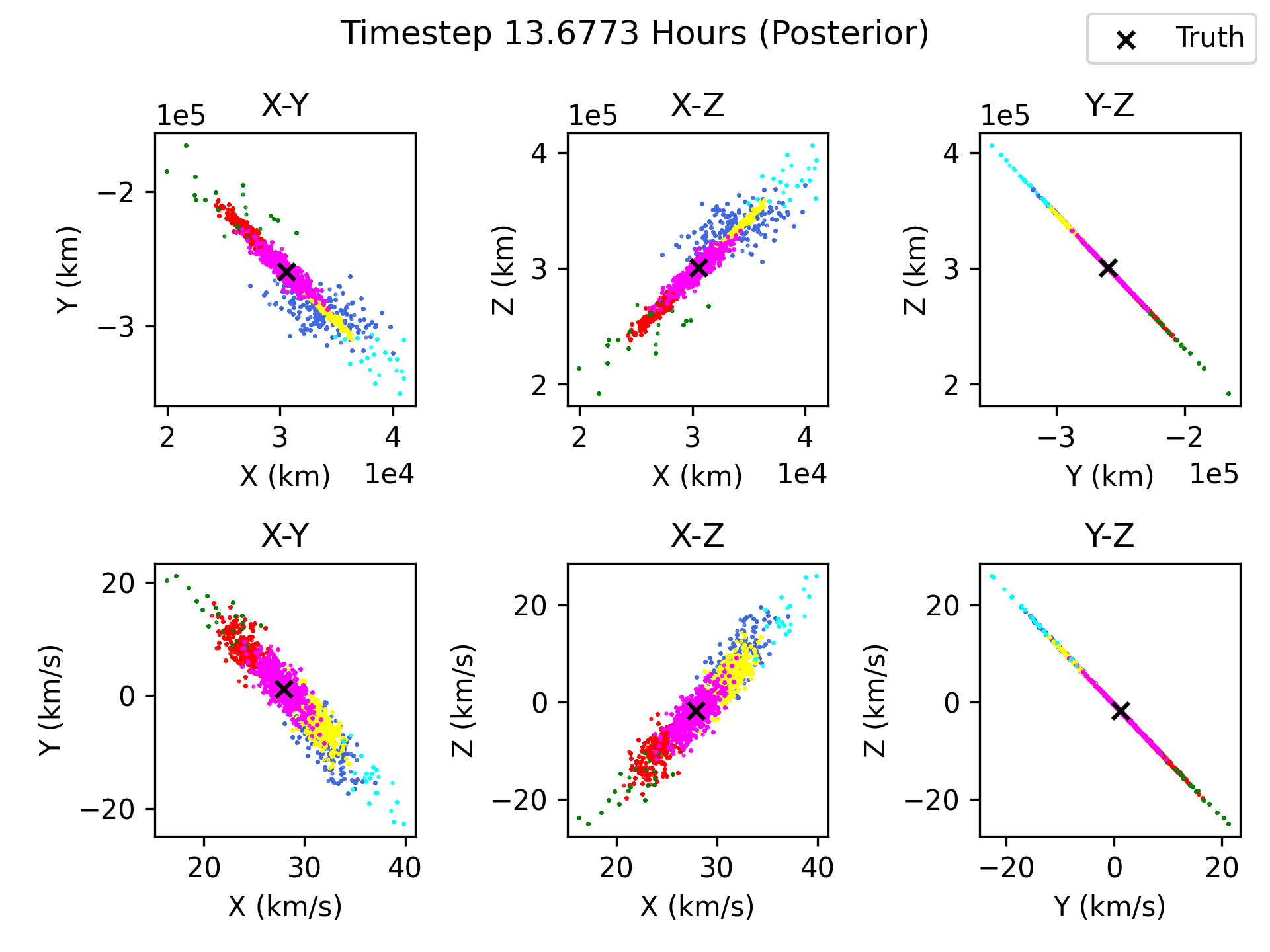}
        \caption{Weight updated ensembles}
        \protect\label{fig:2cUpdate}
    \end{subfigure}
    %\end{comment}

    \caption{State estimate propagation and clustering, MCMC sampling, and ensemble update steps for the PGM-II filter.}
    \label{fig:2pgm2demo}
\end{figure}

To summarize, the PGM-II filter contains the same resampling, propagation, and \textit{a priori} estimate clustering steps as the PGM-I filter. However, the update process of the PGM-II filter is more complex and computationally intensive, involving MCMC sampling, clustering the resulting ensembles, and an ensemble weight update. The steps within the PGM-II algorithm are outlined in Algorithm \ref{alg:pgm2} (copied from Ref. \cite{raihan2018pgm2}). Starting with a clustered \textit{a priori} PDF, the PGM-II update is illustrated by Figure \ref{fig:2pgm2demo}. 

\begin{algorithm}
\caption{Particle Gaussian Mixture Filter II}\label{alg:pgm2}
\begin{algorithmic}
    \State Given $\pi_0 (x) = \sum_{i=1}^{M(0)} \omega_i (0) p_g(x; \mu_i (0), P_i (0))$, transition density kernel $p(x'|x)$, $n = 1$.
    \begin{enumerate}
        \State Sample $N_p$ particles from $\pi_{n-1}$ and the transition density kernel $p_n (x'|x)$ as follows: \label{alg:repeat2}
        \begin{itemize}
            \item Sample $X^{(i)^\prime}$ from $\pi_{n-1} (\cdot)$.
            \item Sample $X^{(i)}$ from $p(\cdot|X^{(i)^\prime})$.
        \end{itemize}
        \State Use a clustering algorithm $\mathcal{C}$ to cluster the set of particles $X^{(i)}$ into $M^{-} (n)$ Gaussian clusters with weights, mean, and covariance given by $\{\omega_i^{-} (n), \mu_i^{-} (n), P_i^{-} (n)\}$.
        \State Use MCMC to sample from the component posteriors $\pi_{i,n}(x)$ to generate the ensembles $A_i$.
        \State Compute the mixture weights $w_i(n)$ by evaluating the sequence of modal likelihoods using the likelihood function definition and Eq. \eqref{eq5a:discreteUpdate}.
        \State Sample N particles from the weighted collection of ensembles $\{w_i(n), A_{n,i}\}$.
        \State $n = n + 1$. Go to Step \ref{alg:repeat2}.
    \end{enumerate}
\end{algorithmic}\end{algorithm}

\section{Hybrid Particle Gaussian Mixture Filtering Method}\label{subsec:ours}
 Most state estimate PDFs may be accurately modeled by a Gaussian Mixture Model, but uninformative priors may require tens, possibly hundreds, of Gaussian mixture components. Indeed, the type of extreme initial state uncertainty which we model in this article may only be represented by a (multivariate) uniform PDF.

\begin{figure}[h!]
	\centering\includegraphics[width=\columnwidth]{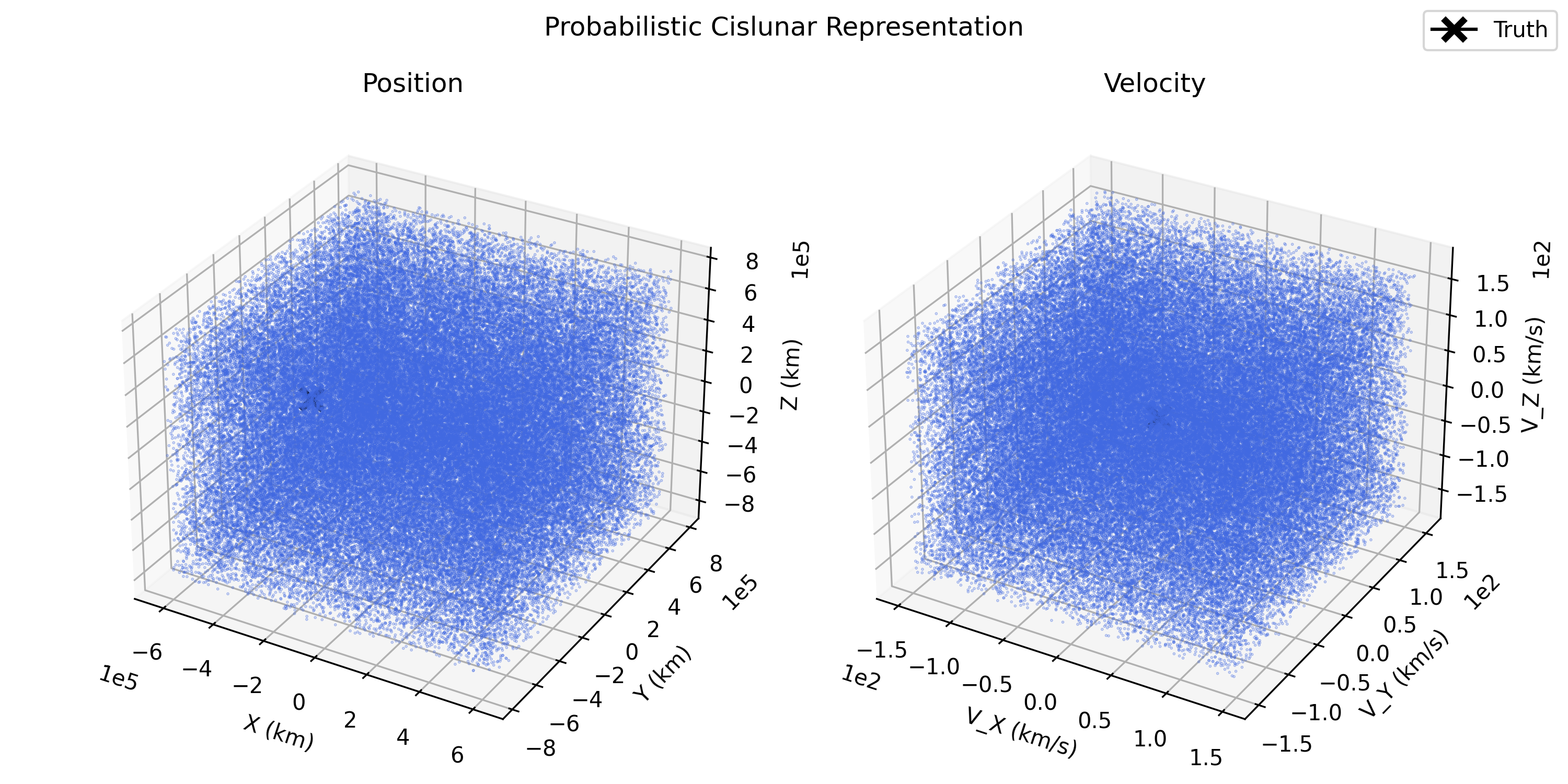}
	\caption{A large multivariate uniform PDF representing $p_0^{-}(\mathbf{x})$ for all examples in Section \ref{sec:Results}}
	\label{fig:2mvUniform}
\end{figure}

An advantage of MCMC sampling is that neither component of the target function in Eq. \eqref{eq9:MCMCsamplingfunc} has to be distributed as Gaussian or as a GMM. To that end, instead of modeling our initial prior as a Gaussian Mixture Model with KF-PGM in Refs. \cite{paranjape2026} and \cite{paranjape2025}, which requires several observations to obtain its initial state estimate, we can model maximum state uncertainty using a multivariate uniform PDF, knowing only that our target is somewhere within the cislunar domain at the time we receive the first observation. In other words, instead of relying on the kinematic fitting based estimate from Refs. \cite{paranjape2026} and \cite{paranjape2025}, we set $p_0^{-}(\mathbf{x})$ as a uniform PDF in the cislunar region (i.e. an \textit{a priori} estimate with minimal information content).  Although $p_0^{-}(\mathbf{x})$ can be expressed by any PDF, our article focuses on the use of a uniform PDF to characterize maximum state uncertainty (even higher than the minimal range assumption based KF-PGM IOD estimate), and how our hybrid PGM-based filter is robust to this extreme initial state uncertainty. Utilizing such a large multivariate uniform at the beginning of our orbit determination process allows us to update with many more measurements than KF-PGM during that critical first pass. Figure \ref{fig:2mvUniform} illustrates this extremely uncertain PDF in the state space. Subsequent state predictions and updates are modeled as GMMs.

% The sampling function $P(\mathbf{x})$ enables us to utilize constraints such as the example in this section by setting $p(\mathbf{z}_k|\mathbf{x}) = p(\mathbf{\theta}|\mathbf{x})$ as a uniform PDF. To differentiate between the measurement model and cislunar target \textit{a priori} information in this paper such as angular rates, range, etc., we use the vector variable $\mathbf{\theta}$ to refer to these non-measurement characteristics. 

The PGM-II algorithm assumes a Gaussian or GMM-based \textit{a priori} estimate fused with a measurement whose noise is distributed as a Gaussian PDF. For this work, we no longer make that assumption. Instead, we modify the MCMC sampling function $P(\mathbf{x})$ for the PGM-II filter such that $p_k^{-}(\mathbf{x})$ can be equated to a highly uncertain, uniform PDF. The resulting PGM-II based update will result in a posterior GMM state estimate. Each subsequent $p_{k}^{-}(\mathbf{x})$ will be modeled as a GMM. Consistent with the notation from Algorithms \ref{alg:pgm1} and \ref{alg:pgm2}, we summarize our hybrid filtering approach in Algorithm \ref{alg:hybrid}. %It is worth noting that $p(\mathbf{z}_k|\mathbf{x})$ does not need to be modeled as Gaussian either. However, those formulations are outside the scope of this work.  %As a result, our hybrid Particle Gaussian Mixture filtering approach considers two types of fusions: 1) Gaussian/GMM-distributed $\mathbf{x}$ with Gaussian-distributed $\mathbf{z}_k$, and 2) uniformly distributed $\mathbf{x}$ with Gaussian/GMM-distributed $\mathbf{z}_k$. The former is only considered for the first iteration of our hybrid filtering technique, while the latter is considered for all subsequent iterations.

\begin{algorithm}
\caption{Hybrid Particle Gaussian Mixture Filtering Method}\label{alg:hybrid}
\begin{algorithmic}
    \State Given a uniformly distributed initial PDF $p_0^{-}(\mathbf{x})$, transition density kernel $p(\mathbf{x}'|\mathbf{x})$, $n = 1$, switching time step $n_k \ge 2$.
    \begin{enumerate}
        \State Sample $N_p$ particles from $\pi_{n-1}$ and the transition density kernel $p_n (x'|x)$ as follows: \label{alg:repeatHybrid}
        \begin{itemize}
            \item Sample $X^{(i)^\prime}$ from $\pi_{n-1} (.)$.
            \item Sample $X^{(i)}$ from $p(.|X^{(i)^\prime})$.
        \end{itemize}
        \State Use a clustering algorithm $\mathcal{C}$ to cluster the set of particles $X^{(i)}$ into $M^{-} (n)$ Gaussian clusters with weights, mean, and covariance given by $\{\omega_i^{-} (n), \mu_i^{-} (n), P_i^{-} (n)\}$.

        \If{$n \leq n_k$}
            \State Follow Steps 3-5 from Algorithm \ref{alg:pgm2}.
        \Else
            \State Follow Step 3 from Algorithm \ref{alg:pgm1}.
        \EndIf

        \State $n = n + 1$. Go to Step \ref{alg:repeatHybrid}.
    \end{enumerate}
\end{algorithmic}\end{algorithm}

Although the MCMC sampling modification in this PGM-II filtering update process is powerful, it is far more computationally intensive than the entirety of the PGM-I update. Thus, it helps to only run the PGM-II based filter for a handful of steps (usually between two to five), switching subsequently to the PGM-I filter for computational efficiency once the uncertainty in the state has been sufficiently reduced. This filtering sequence alludes to the hybrid nature of our filtering technique.

\section{Results and Discussion}\label{sec:Results}

In this section, we present some example results of our hybrid PGM-based filter. We set $p_0^{-}(\mathbf{x})$ equal to the large uniform PDF shown in Figure \ref{fig:2mvUniform}, whose upper and lower bounds for each position and velocity state component are $\pm$550000 km. and $\pm$100 km/s. Although most cislunar RSOs lie well within 550000 km. from the Earth's center and move at speeds lower than 42 km/s if they are not actively escaping the solar system, we set particularly wide bounds to demonstrate by example the robustness of our hybrid PGM-based filter to highly uncertain initial state estimates. By considerably relaxing state vector component bounds relative to most cislunar target tracking literature, we create a minimal assumption framework for robust cislunar target tracking.

\subsection{Hybrid Filtering Approach}\label{subsec:measFuse}

To demonstrate our hybrid PGM-based filtering approach, we utilize the PGM-II based filtering approach for at least the first two time steps to allow for sufficient localization in the velocity space, since a single set of $[AZ, EL]^T$ measurements does not provide any insights about target velocity. Afterwards, we switch to the PGM-I filter. Maintaining the assumption that the measurement noise for $\mathbf{z}_k$ is distributed as zero-mean and Gaussian, we fuse a GMM \textit{a priori} state estimate with a Gaussian-distributed measurement for each time step after the first. For the first time step, we fuse $p_0^{-}(\mathbf{x})$  with the first measurement. Due to intensive computational resources required for MCMC sampling, the PGM-II based update was only performed for the first two sets of measurements to gain some localization in the velocity space.

\begin{figure}[!thpb]
    \centering
    % Subfigure 1: Width of subfigure matches the image width
    \begin{subfigure}{0.8\columnwidth}
        \centering
        \includegraphics[width=\linewidth]{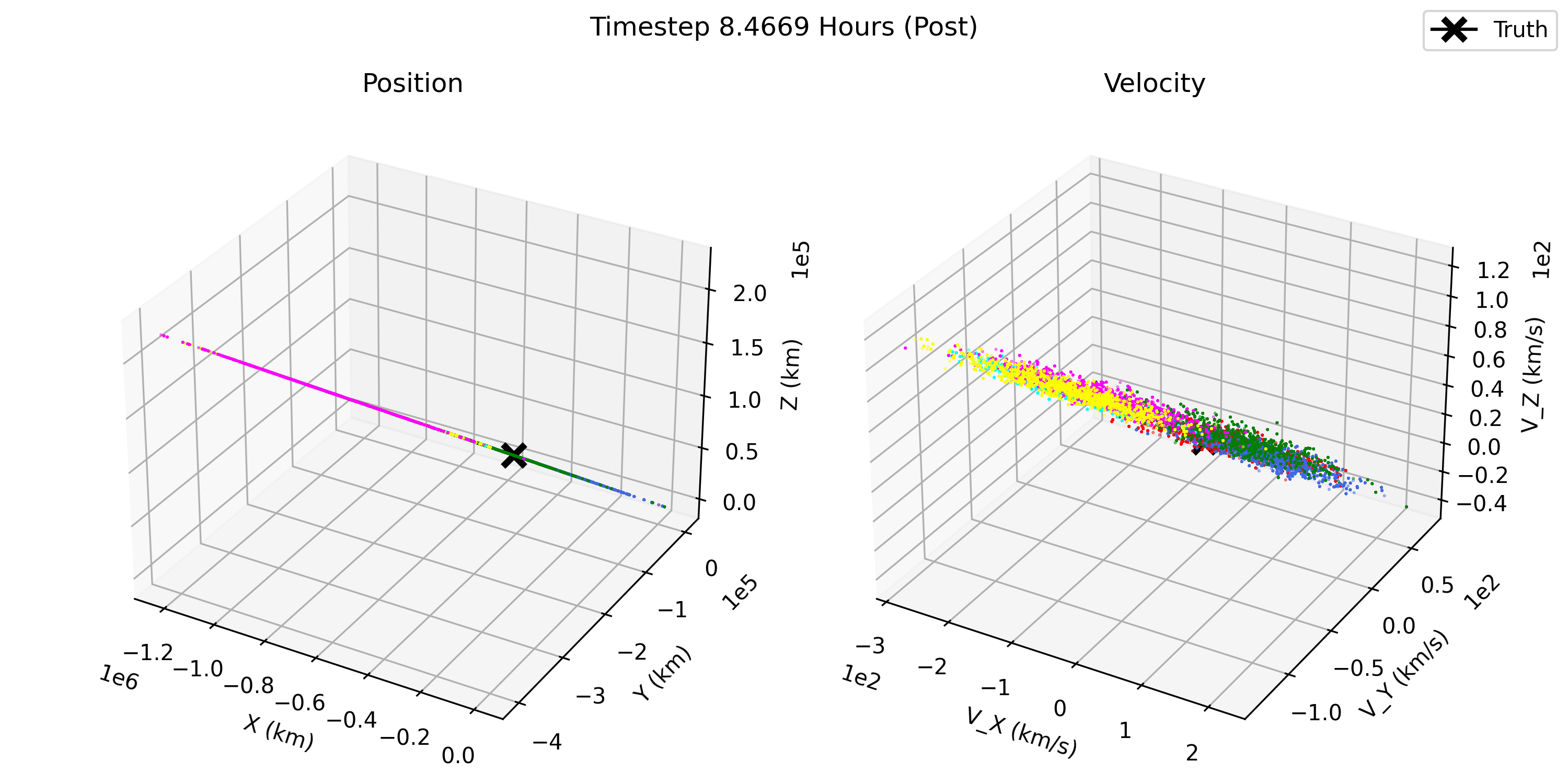}
        \caption{Posterior estimate after two iterations of the PGM-II based measurement update}
        \label{fig:3aPGM2}
    \end{subfigure}

    \vspace{0.2cm} % Optional vertical spacing

    % Subfigure 2: Width of subfigure matches the image width
    \begin{subfigure}{0.8\columnwidth}
        \centering
        \includegraphics[width=\linewidth]{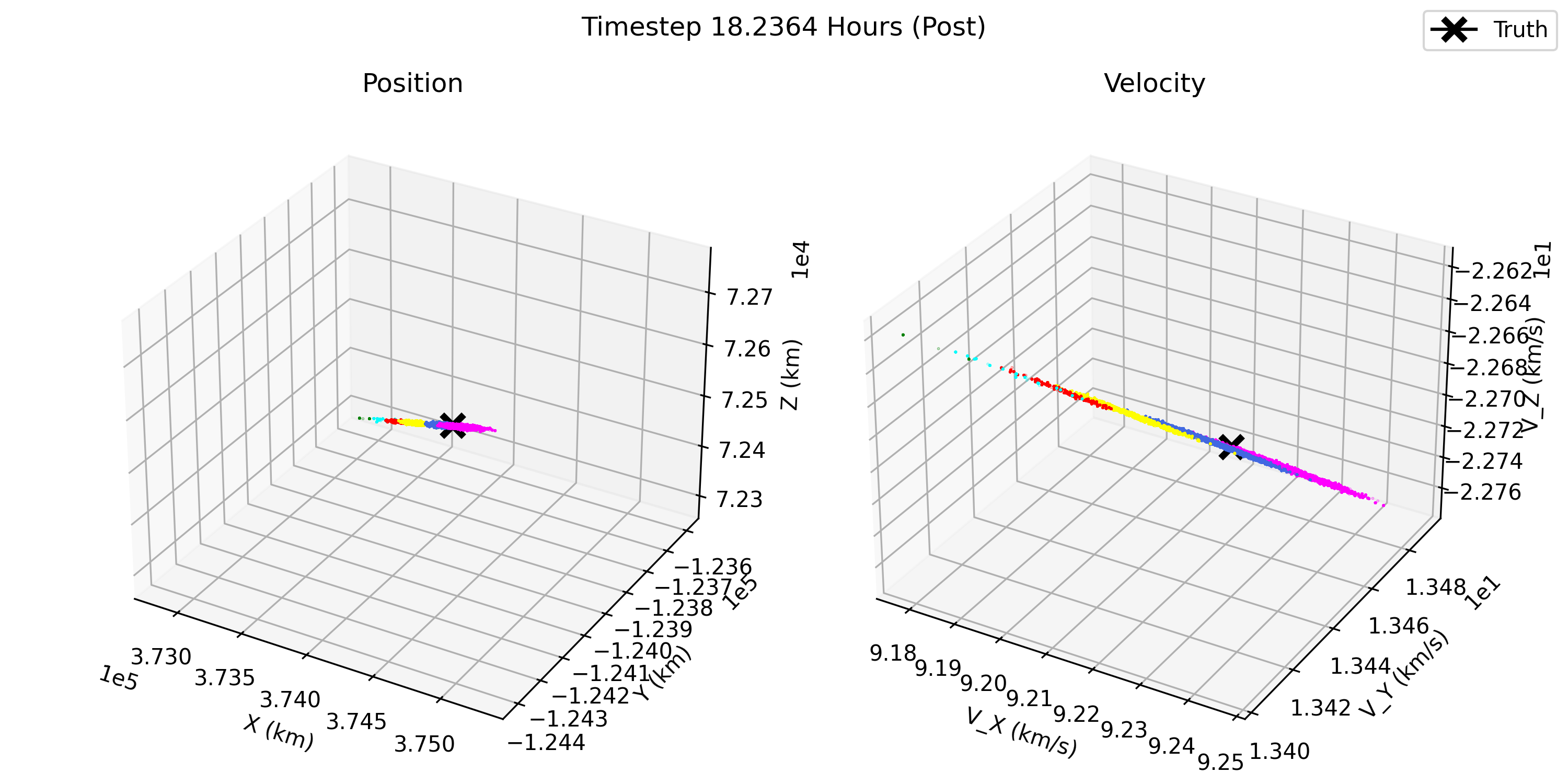}
        \caption{State estimate at the end of the first pass}
        \label{fig:3bHybrid}
    \end{subfigure}

    \caption{Posterior state estimates after two measurements and at the end of the simulation/first pass using our hybrid PGM-based filter.}
    \label{fig:3measOnlyEstimates}
\end{figure}

For all simulations in this section, we showcase a target moving in the aforementioned 9:2 resonant NRHO. This orbit was chosen due to its stable, elliptic motion about the Moon and its importance for NASA's Gateway and future Artemis missions. Treating the cislunar RSO as a new target of interest, our College Station-based observer is tasked to take measurements of the target roughly every 40 minutes before the target disappears from our sensor field of view, at which point the simulation ends. This measurement schedule is identical to that of the first pass of several examples in Refs. \cite{paranjape2026} and \cite{paranjape2025}. Starting with the initial state estimate shown in Figure \ref{fig:2mvUniform}, we show the posterior estimates $p_k^{+}(\mathbf{x})$ at the end of the second time step -- when we switch from a PGM-II based update to a PGM-I update -- and at the end of the first pass in Figure \ref{fig:3measOnlyEstimates}.

\subsection{Comparison to PGM-I Based Filtering Techniques}
\begin{figure}[!thpb]
    \centering
    % Subfigure 1: Width of subfigure matches the image width
    \begin{subfigure}{0.8\columnwidth}
        \centering
        \includegraphics[width=\linewidth]{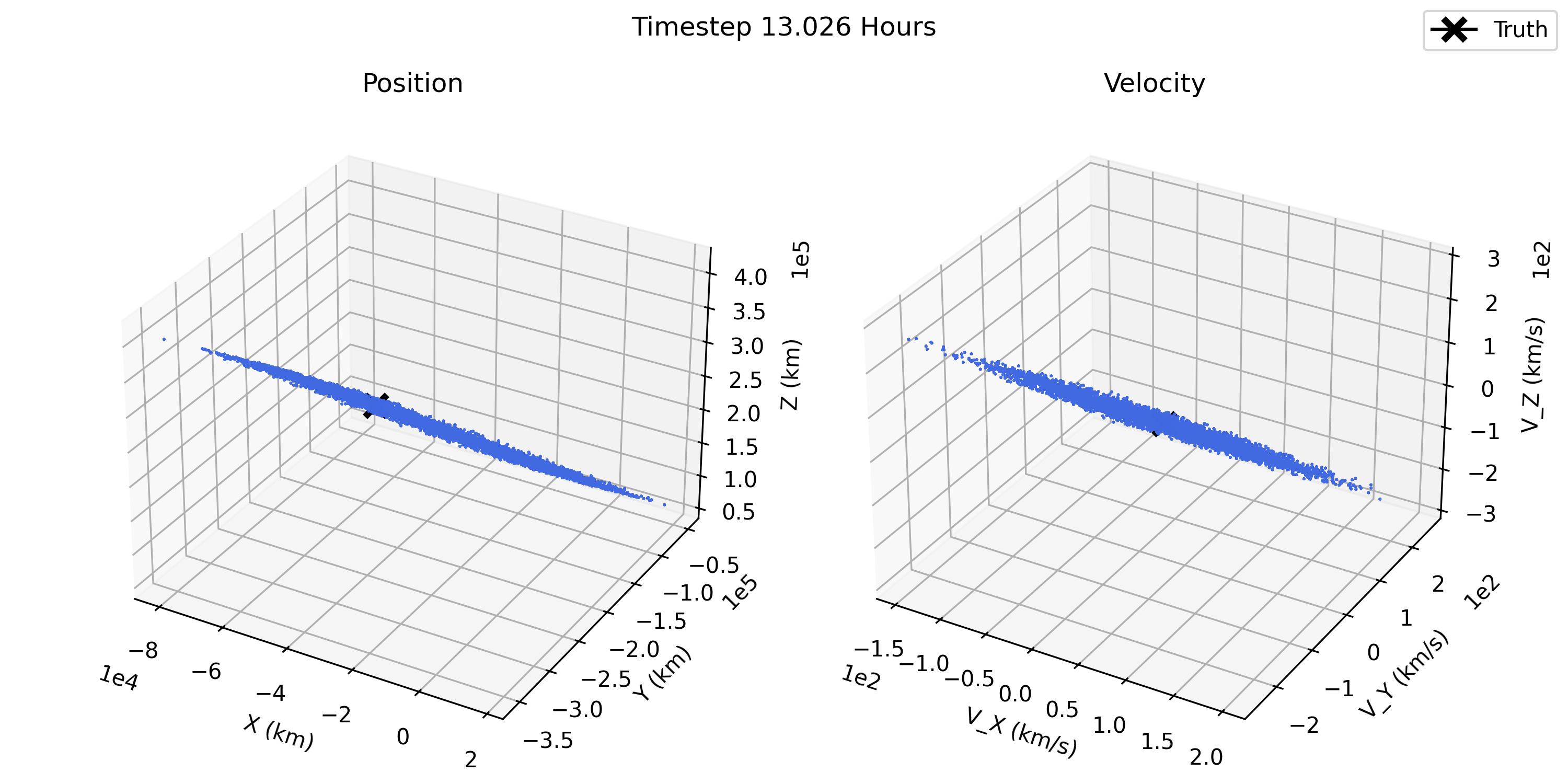}
        \caption{KF-PGM initial state estimate}
        \label{fig:5aIOD}
    \end{subfigure}

    \vspace{0.2cm} % Optional vertical spacing

    % Subfigure 2: Width of subfigure matches the image width
    \begin{subfigure}{0.8\columnwidth}
        \centering
        \includegraphics[width=\linewidth]{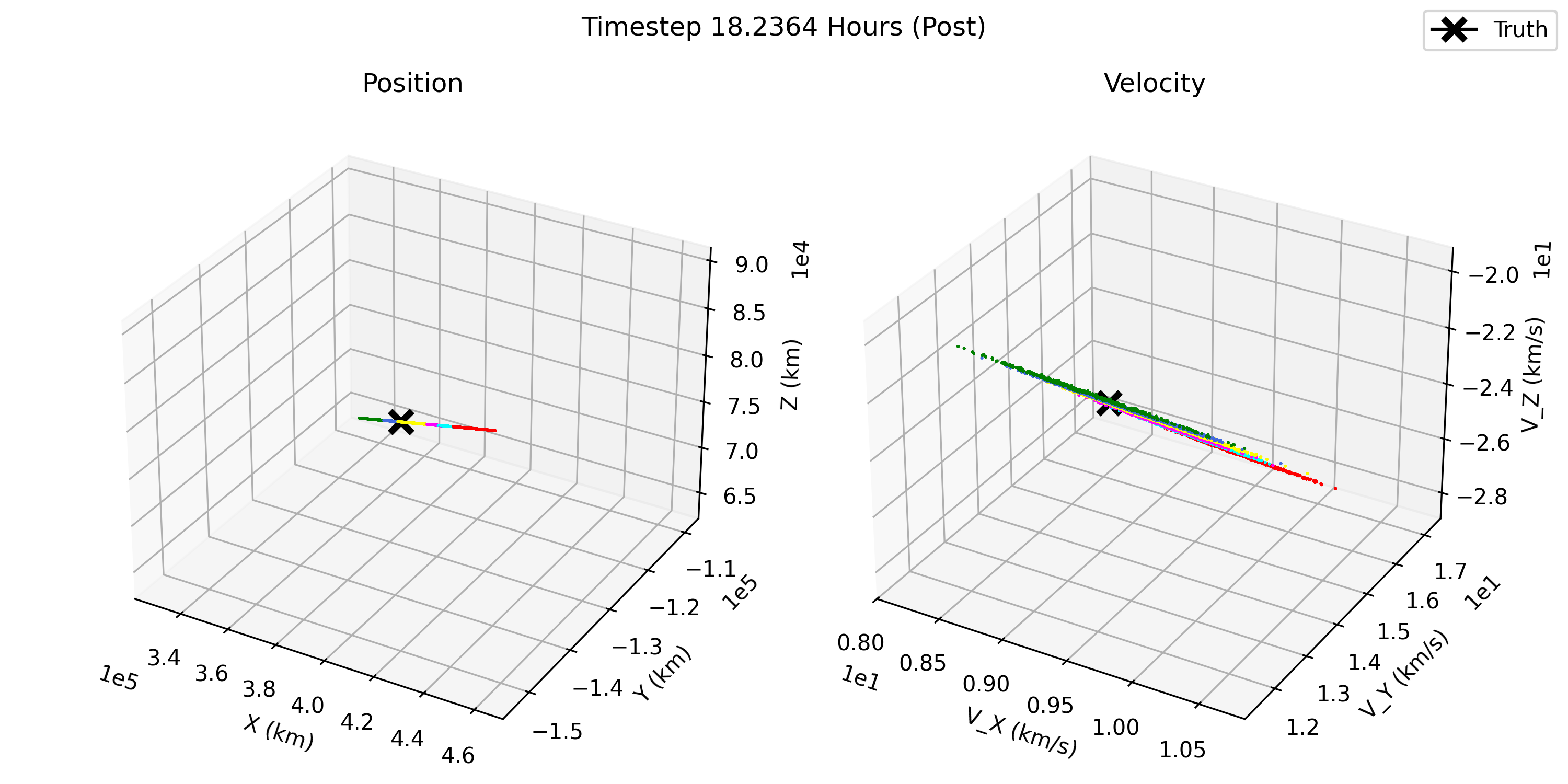}
        \caption{State estimate at the end of a single pass with the KF-PGM framework}
        \label{fig5b:final}
    \end{subfigure}

    \caption{Single pass filtering results for a target within the 9:2 NRHO orbit}
    \label{fig:5kfPGMdemo}
\end{figure}

\begin{figure}[!thpb]
    \centering
    % Subfigure 1: Width of subfigure matches the image width
    \begin{subfigure}{0.8\columnwidth}
        \centering
        \includegraphics[width=\linewidth]{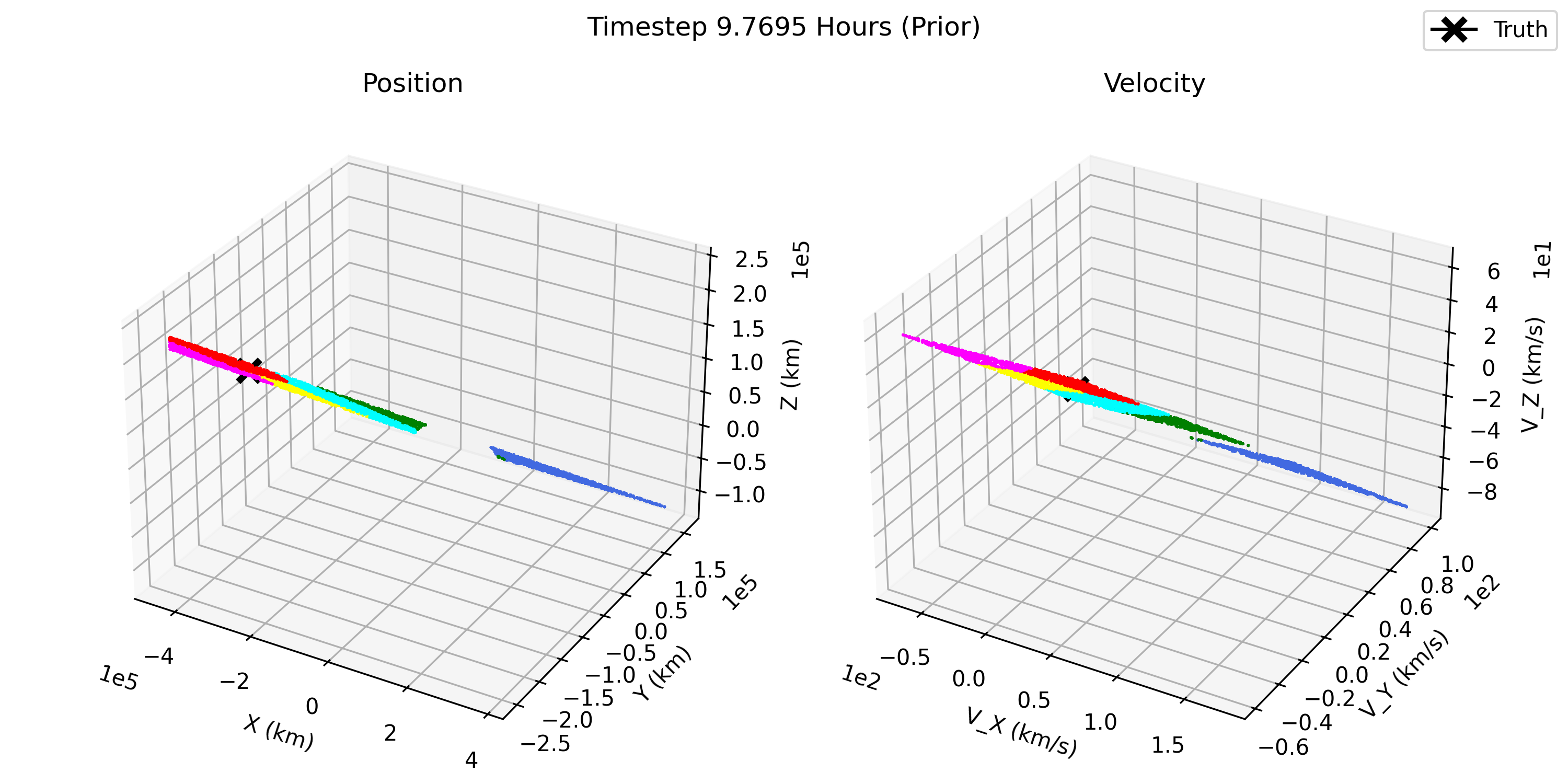}
        \caption{\textit{A priori} estimate propagated and clustered from an IOD estimate formulated from a linear kinematic fit}
        \label{fig:4aPrior}
    \end{subfigure}

    \vspace{0.2cm} % Optional vertical spacing

    % Subfigure 2: Width of subfigure matches the image width
    \begin{subfigure}{0.8\columnwidth}
        \centering
        \includegraphics[width=\linewidth]{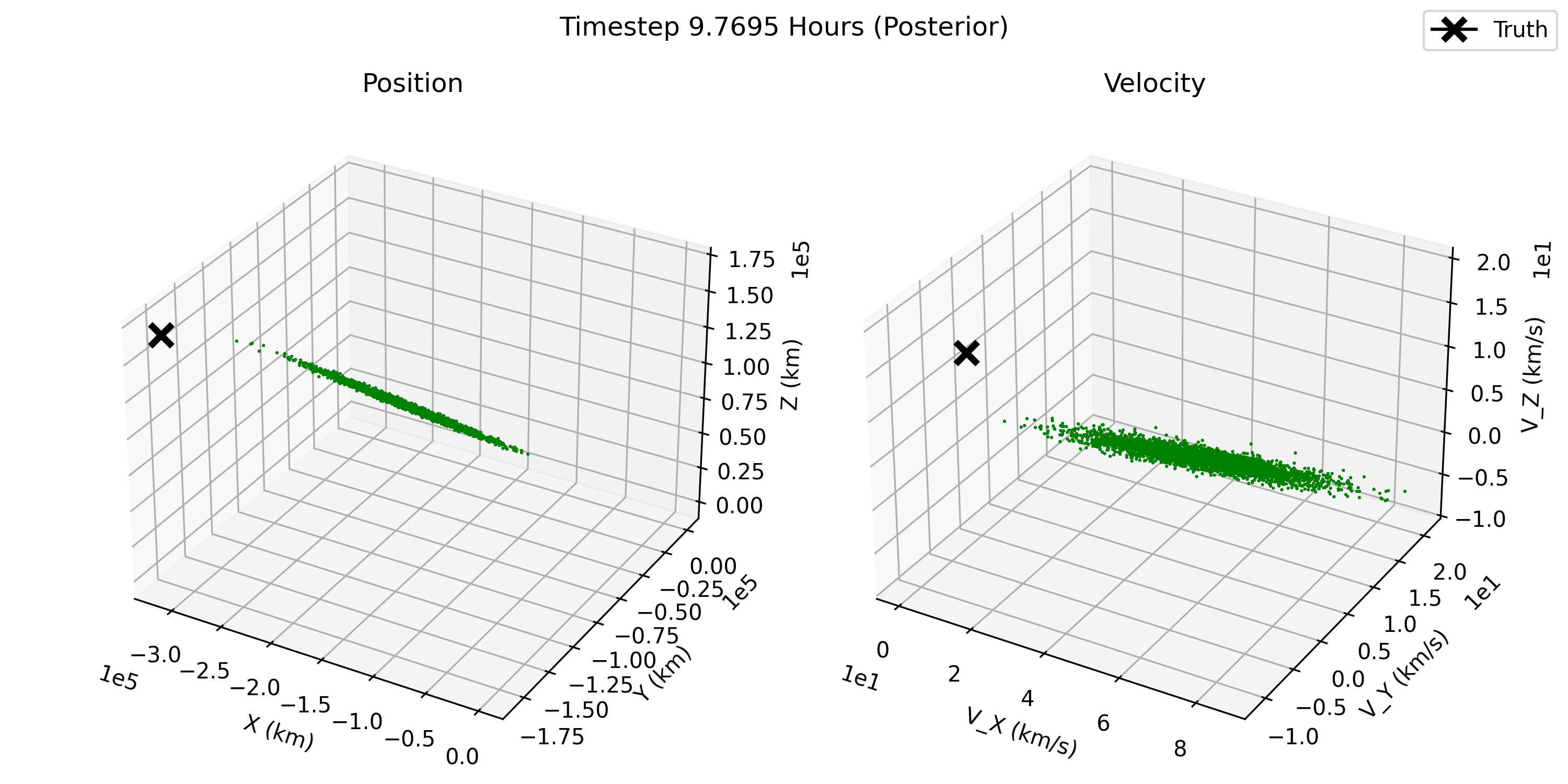}
        \caption{An inconsistent posterior estimate}
        \label{fig:4bPosterior}
    \end{subfigure}

    \caption{Limitations of utilizing KF-PGM with a linear fit between the first two measurements}
    \label{fig:4pgm1fail}
\end{figure}

The remainder of this section focuses on comparison of our hybrid filtering technique to KF-PGM. The KF-PGM framework generates a probabilistic initial state estimate by fusing several consecutive $[AZ, EL]^T$ measurements with a large, loosely defined PDF of the range, which yields an initial position estimate which we can estimate strictly as a function of time with polynomials or splines. This kinematic fit of the position is derived with respect to time to yield an initial velocity estimate. Together, these fits yield an initial state estimate at the time step of the last observation through which the polynomial is fit. This process is repeated several thousands of times by perturbing $[AZ, EL]^T$ measurements with measurement noise and drawing several i.i.d. samples of the range information at each observation time step. The result is an IOD estimate in particle cloud form shown in Figure \ref{fig:5aIOD}. Subsequently, we utilize the PGM-I filter to yield a localized state estimate at the end of the first pass, given by Figure \ref{fig5b:final}.
Another reason we choose to utilize a PGM-II based update for the first two measurement sets is to demonstrate how well this hybrid PGM filtering technique outperforms the KF-PGM framework. Theoretically, the kinematic fitting framework of KF-PGM requires a minimum of two measurements for its lowest order kinematic fit (a line). If the kinematic fit is performed through these first two points to obtain an initial state estimate, this IOD estimate will be consistent with the target truth. However, the PGM-I filter will fail at the subsequent timestep's update step, as demonstrated by Figure \ref{fig:4pgm1fail}. Thus, our proposed hybrid filtering approach illustrates how much fewer observations are really required.

\begin{figure}[!thpb]
    \centering
    % Subfigure 1: Width of subfigure matches the image width
    \begin{subfigure}{0.8\columnwidth}
        \centering
        \includegraphics[width=\linewidth]{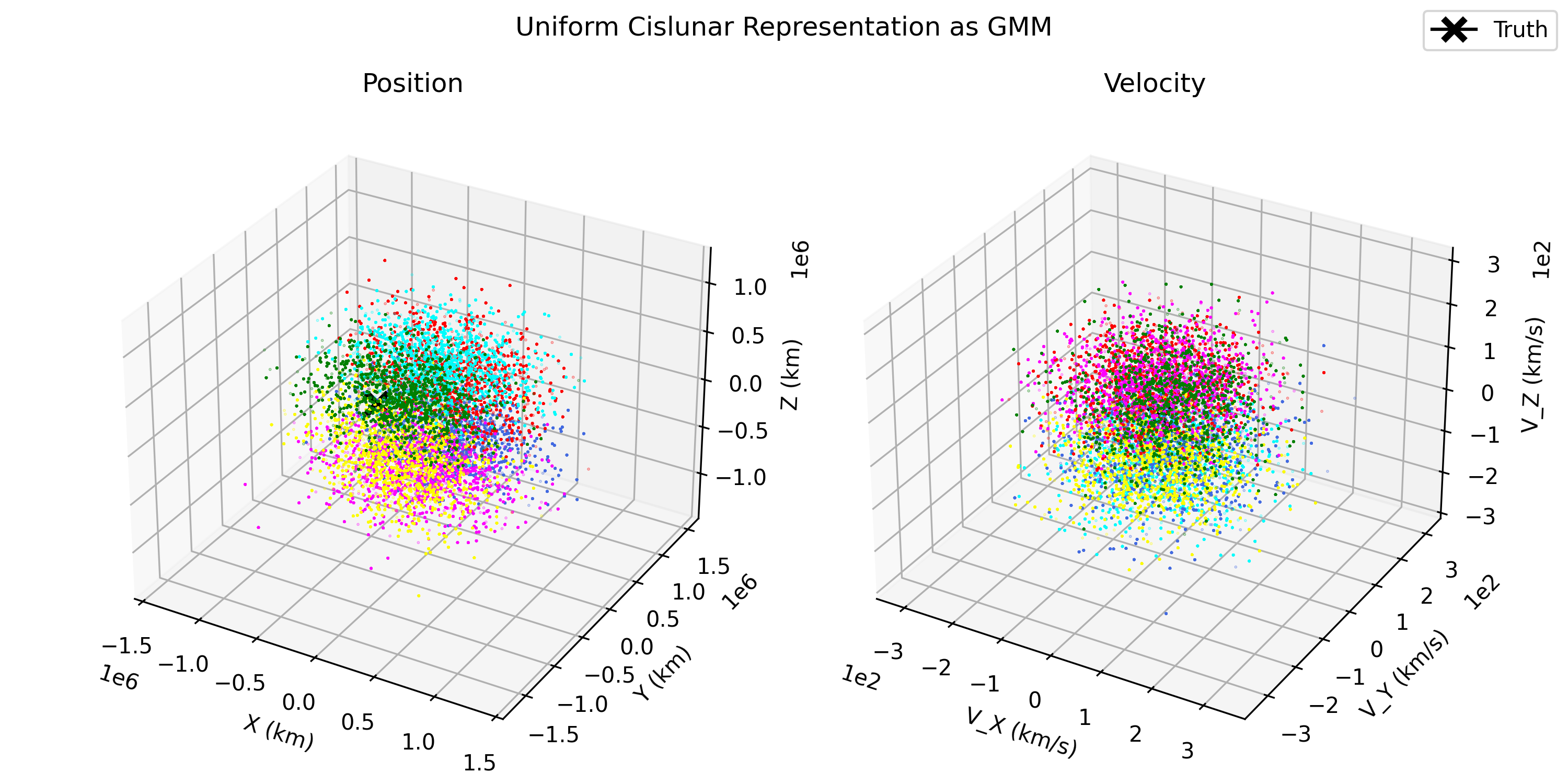}
        \caption{Initial state estimate $p_0^{-}(\mathbf{x})$ resulting from clustering the uniform PDF in Figure \ref{fig:2mvUniform} into a GMM}
        \label{fig:4aPrior}
    \end{subfigure}

    \vspace{0.2cm} % Optional vertical spacing

    % Subfigure 2: Width of subfigure matches the image width
    \begin{subfigure}{0.8\columnwidth}
        \centering
        \includegraphics[width=\linewidth]{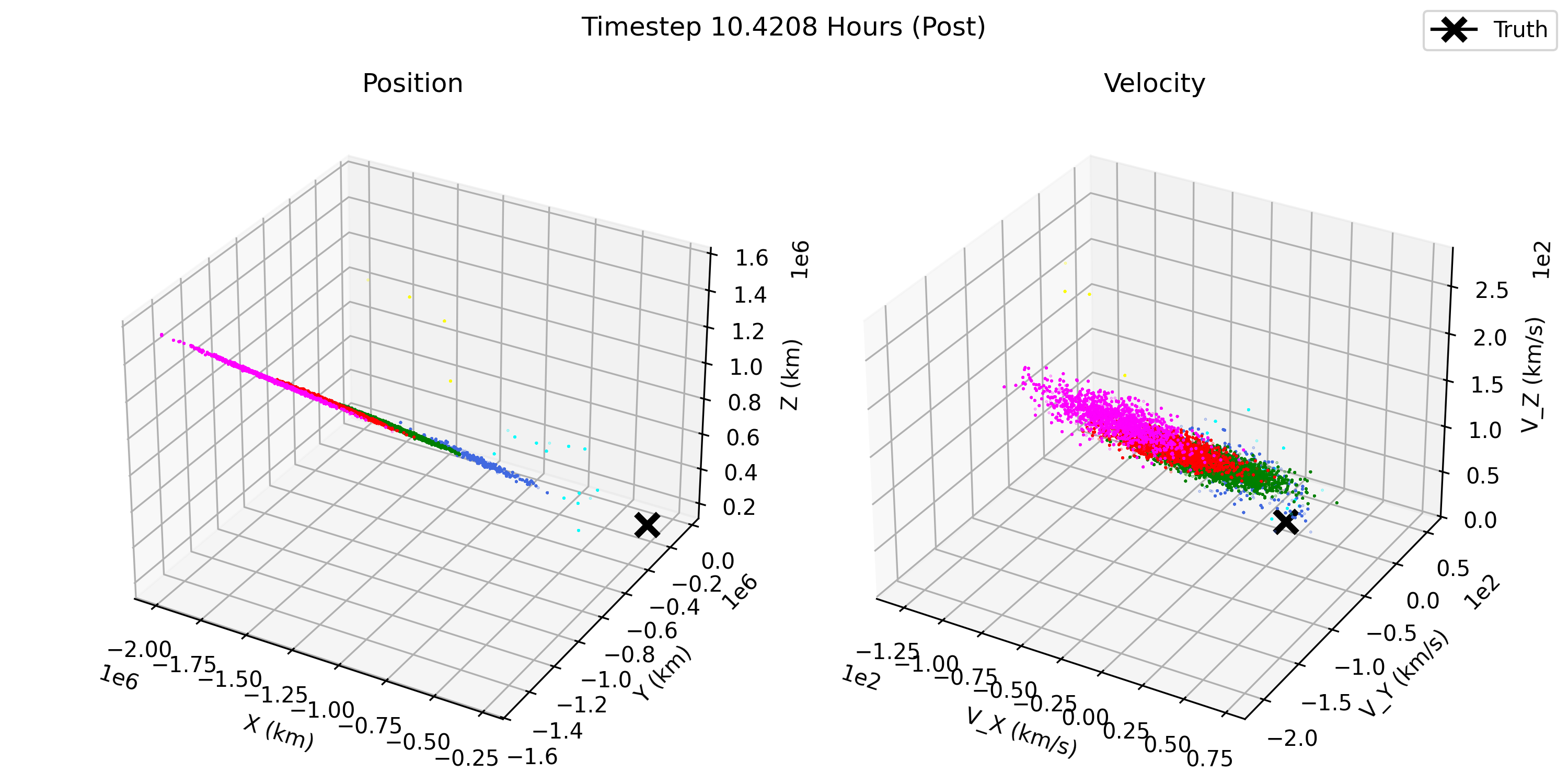}
        \caption{An inconsistent posterior estimate after four update iterations}
        \label{fig:4bPosterior}
    \end{subfigure}

    \caption{Limitations of utilizing solely a PGM-I filtering approach throughout the first pass}
    \label{fig:6pgm1fail}
\end{figure}

To demonstrate that the PGM-I filter alone is insufficient for orbit determination within this first pass of the 9:2 resonant NRHO based target, we cluster all of the blue particles in the multivariate uniform PDF in Figure \ref{fig:2mvUniform} into six clusters to develop a GMM-based \textit{a priori} estimate shown in Figure \ref{fig:6pgm1fail}. Although the PGM-I filter yields a consistent estimate after the first three update cycles starting with this clustered PDF, the resulting entropy tends to be much higher than the estimate resulting from a PGM-II based update. More importantly, after the fourth update step, the PGM-I update loses custody of the target, whereas the hybrid PGM filtering technique in Figure \ref{fig:3measOnlyEstimates} stays consistent through the end of the first pass and simulation.

To compare the filter performances and precision of our hybrid PGM-based filter and the PGM-I filter from the first 9:2 NRHO target tracking example from Ref. \cite{paranjape2026}, we utilize a particle-based entropy metric, given by $H(k) = \mathbb{E}[- \log{p_{k}(\mathbf{x}_i)}] \approx -\frac{1}{N} \sum_{i=1}^{N} \log{p_k(\mathbf{x}_i)}$, where $\mathbf{x}_i$ represents the $i$-th particle in the posterior state estimate. For direct and visual comparisons, we provide plots of the initial state estimate as well as the posterior estimate at the end of the first pass using the KF-PGM framework in Figure \ref{fig:5kfPGMdemo}. 

\begin{figure}[thpb] 
      \centering
      \includegraphics[scale=0.45]{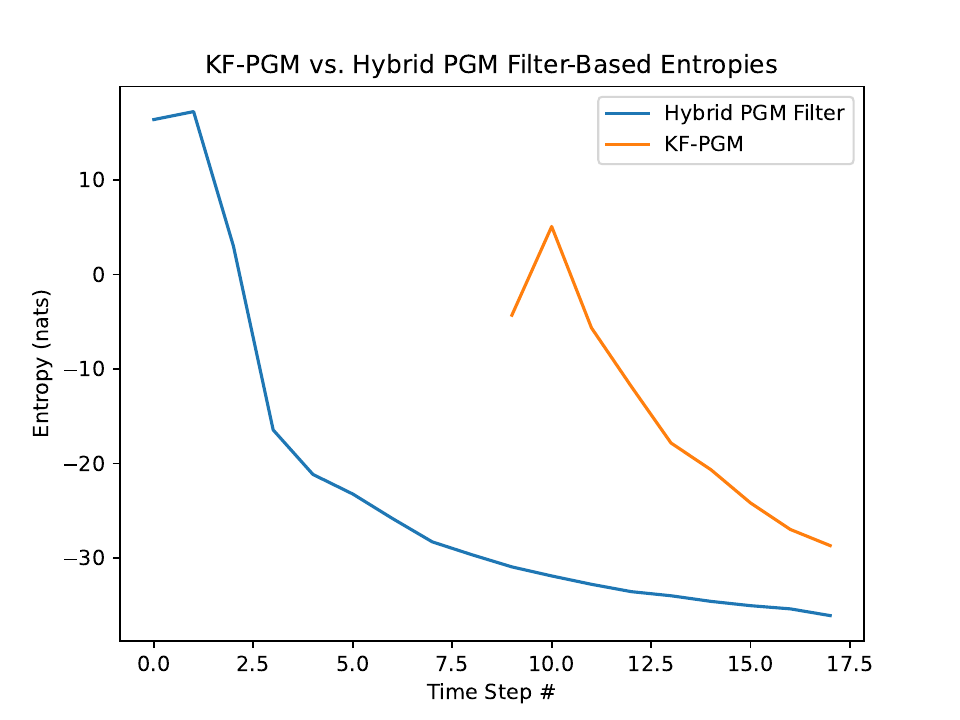}
      \caption{Comparison of entropies during the first pass of a cislunar RSO in the 9:2 NRHO orbit}
      \label{fig7:secAcomp}
\end{figure}

A quick glance between Figures \ref{fig:5kfPGMdemo} and \ref{fig:3bHybrid} show that our hybrid PGM-based filter better localizes the state estimate at the end of the first 9:2 NRHO pass by a factor between 10 to 100 for \textit{each} state vector component! Part of this may be explained by the fact that our hybrid PGM filtering method does not require use of any observations for generating an initial state estimate. Furthermore, the powerful MCMC sampling step of the PGM-II filter enables us to sufficiently sample regions of high probability density consistent with our observations. We formalize these findings by a comparison of entropies between KF-PGM and our hybrid PGM-based filter in Figure \ref{fig7:secAcomp}, and a table of standard deviation reductions in Table \ref{table:stDevs}.

\begin{table}[h]
    \caption{Comparison of state estimate standard deviations between KF-PGM and our hybrid technique from the initial state estimate to the end of the first pass}
    \label{table:stDevs}
    \centering % Use \centering instead of the center environment
    % Use tabular* to span the width of the column.
    % The @{\extracolsep{\fill}} automatically adds space between columns.
    \begin{tabularx}{\columnwidth}{l C C C C} 
    \toprule
    Component & KF-PGM Start & KF-PGM End & Hybrid Start & Hybrid End \\
    \hline
    \midrule % Use \midrule for internal rules with booktabs
    $x$ (km)      & 15,370   & 29,869   & 359,028 & 406.47   \\
    $y$ (km)      & 82,919   & 9,902.4  & 444,363 & 135.18   \\
    $z$ (km)      & 95,764   & 5,791.4  & 449,691 & 77.709   \\
    $\dot{x}$ (km/s) & 54.047  & 0.32216 & 85.836 & 0.010630 \\
    $\dot{y}$ (km/s) & 78.993  & 1.2984  & 88.528 & 0.014715 \\
    $\dot{z}$ (km/s) & 88.644  & 1.9414  & 97.135 & 0.025009 \\
    \bottomrule
    \end{tabularx}
\end{table}

Noticeably, there is a slight upward tick in the entropy during the first 1-2 steps of both KF-PGM as well as our hybrid PGM filter. This is due to high uncertainty and explosive growth in the velocity space in the absence of multiple measurements. This observation further emphasizes the need to utilize the PGM-II filter at least twice before switching to the PGM-I filter.

\section{Conclusions and Future Work}\label{sec:Conclusions}

In this article, we developed and demonstrated a hybrid Particle Gaussian Mixture filter-based orbit determination framework which slightly relaxes an assumption in the PGM-II filter's MCMC sampling step. Even though this filter starts with minimal information about a cislunar RSO's state, its localization can be up to 100 times better than that of the KF-PGM framework in the short term over the same set of observations. We also demonstrated some of the shortcomings of the KF-PGM technique in the event of limited or fewer observations, as well as why it is important to utilize a PGM-II based update step before switching to the PGM-I based update step. This work shows that cislunar target tracking can be done purely in a recursive filtering fashion without the need for targeted cislunar IOD approaches -- which is difficult owing to the invalidity of the two body assumptions central to Gauss's method -- and still obtain highly localized estimates of the target within one pass.

We hope to build upon this work in three ways. First, we hope to incorporate fusion of cislunar RSO characteristics such as range, orbit type, energy, and Jacobi's constant by abstracting these quantities into PDFs. Second, we hope to extend our hybrid PGM-based filter to robustly estimate cislunar RSOs in the presence of non-Gaussian measurement noise. Finally, we hope to extend this hybrid filtering approach to multi-target tracking scenarios in a cislunar environment, where minimal \textit{a priori} knowledge of target states is available. 

\addtolength{\textheight}{-12cm}   % This command serves to balance the column lengths
                                  % on the last page of the document manually. It shortens
                                  % the textheight of the last page by a suitable amount.
                                  % This command does not take effect until the next page
                                  % so it should come on the page before the last. Make
                                  % sure that you do not shorten the textheight too much.

%%%%%%%%%%%%%%%%%%%%%%%%%%%%%%%%%%%%%%%%%%%%%%%%%%%%%%%%%%%%%%%%%%%%%%%%%%%%%%%%

%%%%%%%%%%%%%%%%%%%%%%%%%%%%%%%%%%%%%%%%%%%%%%%%%%%%%%%%%%%%%%%%%%%%%%%%%%%%%%%%

%%%%%%%%%%%%%%%%%%%%%%%%%%%%%%%%%%%%%%%%%%%%%%%%%%%%%%%%%%%%%%%%%%%%%%%%%%%%%%%%
% \section*{APPENDIX}

% Appendixes should appear before the acknowledgment.

%%%%%%%%%%%%%%%%%%%%%%%%%%%%%%%%%%%%%%%%%%%%%%%%%%%%%%%%%%%%%%%%%%%%%%%%%%%%%%%%

%References are important to the reader; therefore, each citation must be complete and correct. If at all possible, references should be commonly available publications.

\end{document}